# Modulating the electronic properties of graphdiyne nanoribbons

Zheng-Zhe Lin*, Qun Wei and Xuanmin Zhu

*School of Physics and Optoelectronic Engineering, Xidian University, Xi'an 710071, China*

**Abstract** - By *ab initio* calculation, Au, Cu, Fe, Ni and Pt adatoms were proposed for modulating the electronic property of graphdiyne naoribbons (GDNRs). GDNRs of 1~4 nm in width were found to be stable at room temperature, and the thermal rates of Au, Cu, Fe, Ni and Pt adatoms escaping from GDNR are slower than 0.003 atoms per hour even at 900 K. According to the calculation, Au and Cu-decorated GDNRs are metallic with carrier concentrations close to that of graphene at room temperature, while Fe, Ni and Pt-decorated GDNRs are n-type semiconductors with impurity states below Fermi energy. Heterojunction composed by doping Au, Cu or Fe atom on one side of GDNR was proposed as metal-semiconductor rectifier with rectification ratio of 2.8, 1.5 or 2.5 at 1.0 V, respectively.

## 1. Introduction

As the element with the most allotropes in nature, C was proposed as a candidate of material for emerging electronics. In the past decades, C nanotubes have been widely used as the material for building nanodevices. Recently, single-layer graphene was successfully prepared, and with remarkable electronic properties it was proposed to be a good material to build next-generation integrated circuits [1, 2]. However, usual two- or three-dimensional C materials are not semiconductors, e.g. graphite or graphene is metallic and diamond is insulator with a wide band gap, and the techniques for modulating the band gap have to be used in building transistors or other devices. Although graphene nanoribbon with armchair edge has a width-dependent band gap [3, 4], complicated techniques should be implemented for fabricating graphene nanoribbons with uniform width [5]. Even though, edge reconstruction of graphene nanoribbons may spontaneously happen [6, 7] and has an influence on electronic transport [8]. More than 20 years ago, graphyne, a hypothetical graphite-like layered C allotrope composed of $sp^2$ and $sp$ C atoms, was theoretically predicted to be semiconductor [9, 10]. Since long time ago, graphyne and its family, named graphdiyne, graphyne-3 *et al*. [11], have not been experimentally prepared. Recently, large area graphdiyne films were synthesized on the surface of Cu by a cross-coupling reaction using hexaethynylbenzene [12], and a technique was developed for growing graphdiyne nanowires [13]. Graphdiyne may be the most stable in artificially synthesized C allotropes [14], and has shown an improved performance in polymer solar cells [15]. Recent theoretical studies mainly concentrate on the basic electronic properties of graphdiyne [16, 17] and other hypothetical graphyne-like structures [18, 19]. To design graphdiyne-based devices, theoretical research on modulating the electronic properties of graphdiyne should be beneficial to guild corresponding experiments.

---

* Corresponding author. Tel.: Fax: +86 81891371. E-mail address: linzhengzhe@hotmail.com (Z.-Z. Lin).

In this work, *ab initio* calculations were performed for Au, Cu, Fe, Ni and Pt adatoms on graphdiyne to study the modulation of electronic properties. By investigating the rate of thermal bond reconstruction, it was indicated that graphdiyne naoribbons (GDNRs) are stable at room temperature. With large adsorption energy, Au, Cu, Fe, Ni and Pt adatoms on GDNR have very slow escaping rates from the surface even at 900 K. By these metal adatoms, Fermi surface of GDNR was lifted without severely changing the energy band profile. By 0.5% doping ratio, at room temperature the carrier concentrations of Au and Cu-decorated GDNRs get close to that of graphene, while Fe, Ni and Pt-decorated GDNRs are n-type semiconductors with impurity states below Fermi energy. By controlling the deposition rate of specific metal atom on graphdiyne, the electronic property of metallic or semiconducting GDNR could be artificially controlled. On this basis, the possible application of heterojunction composed by doping metal atoms on one side of GDNR as metal-semiconductor rectifier was studied.

## 2. Theory

To investigate the electronic property of graphdiyne and corresponding structures,

density functional theory (DFT) calculations were performed using the SIESTA code [20]. The generalized gradient approximation (GGA) of Perdew-Burke-Ernzerhof [21] and Troullier-Martins norm-conserving pseudopotential [22] were adopted. The grid mesh cutoff was set 150 Ry. For structure optimization and energy band calculation, double-zeta-polarized basis set and a 20 Å vacuum layer was used. The structures were relaxed until the atomic forces were less than 0.01 eV/Å. Nudged elastic band method [23-25] was applied to determine the barriers and the minimum-energy paths (MEPs) from one structure transforming into another.

For quantum transport, calculations were performed using non-equilibrium Green's function method [26] as implemented in the TRANSIESTA module [27] of SIESTA. The electrode calculations were performed under periodical boundary conditions. For a bias voltage $V_b$ applied on the system, the current is given by Landauer-Buttiker formula [28]

$$I = \frac{2e}{h} \int T(E,V_b)[f_L(E-E_F-\frac{eV_b}{2}) - f_R(E-E_F+\frac{eV_b}{2})]dE, \qquad (1)$$

where $T(E, V_b)$ the transmission rate of the band state at energy $E$, $E_F$ the Fermi energy of the electrodes, and $f_L$ and $f_R$ the Fermi-Dirac distribution functions of both electrodes. To save computation time, single-zeta-polarized basis set was applied for C atoms, while double-zeta-polarized basis set was applied for other elements.

## 3. Results and discussion

3.1 *Basic properties*

Before exploring the electronic property, the structure, energy band and thermal stability of graphdiyne were investigated. The graphdiyne sheet is composed by hexagonal rings of *sp²* C atoms connected by 4-atom *sp* C-C chains. The structure of single-layer graphdiyne is shown in upper Fig. 1(a), where the rhombic unit cell is drawn by solid line. The optimized lattice constant was $a_0$=9.50 Å, in good agreement with the previous value of 9.48 Å [16] calculated by the projector-augmented-wave method. The Brillouin zone and energy band calculated by 10×10×1 *k*-point sampling are shown in lower Fig. 1(a). Graphdiyne sheet is semiconductor with direct band gap of 0.49 eV at Γ-point, which is close to the previous value of 0.46 eV [16]. It should be noted that the band gap is always underestimated by the DFT calculations.

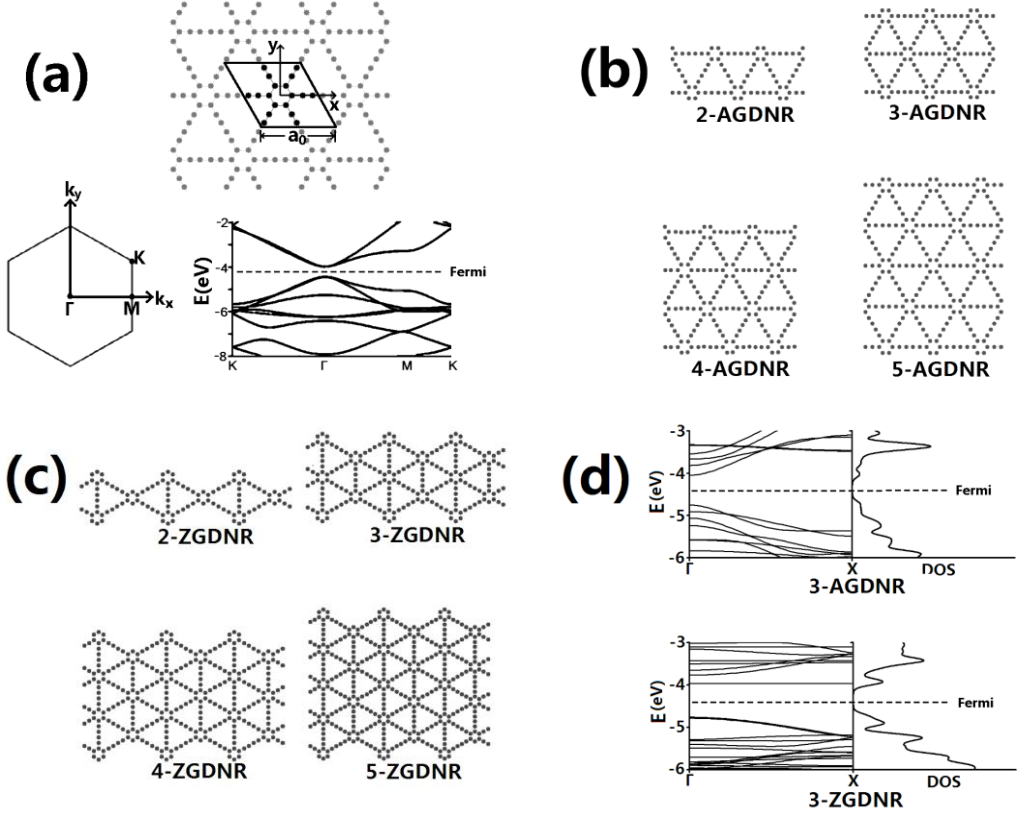

Fig. 1 (a) the structure of single-layer graphdiyne with unit cell shown by solid line (upper), the corresponding Brillouin zone (lower left) and energy band (lower right). The structures of AGDNRs and ZGDNRs are shown in (b) and (c), respectively. For 3-AGDNR and 3-ZGDNR, the energy bands and DOS are shown in the upper and lower panel of (d), where Γ the center and X the boundary of the Brillouin zone.

The structures of GDNR with an armchair (AGDNR) or zigzag edge (ZGDNR) are shown in Fig. 1(b) and (c), respectively, denoted as *n*-AGDNR and *n*-ZGDNR. For AGDNRs only nonmagnetic states were found, while ZGDNRs can be in nonmagnetic or ferromagnetic states. Very small geometry difference was found between ferromagnetic and nonmagnetic ZGDNRs. For one unit cell of 2~5-ZGDNR, the energy of ferromagnetic state is about 30 meV lower than nonmagnetic state. The difference between the distance *a* of neighboring hexagonal C rings in GDNR and $a_0$ of graphdiyne is less than 2%, and no obvious deformation was found on GDNR edges. The energy band was calculated by $100 \times 1 \times 1$ $k$-point sampling and direct band gap at Γ-point was found for both AGDNR and ZGDNR. As an example, the energy bands and density of states (DOS) of 3-AGDNR and 3-ZGDNR are plotted in Fig. 1(d). The band gap of AGDNR and ZGDNR decreases and approaches the value of graphdiyne sheet with increasing width. For 2~5-AGDNR and ZGDNR, the band gaps are in the range of 0.79~0.61 and 1.18~0.62 eV, respectively. Localized edge states of π* orbitals were found above the Fermi level of ZGDNR, presenting an almost straight $E$~$k$ line, e.g. seeing lower Fig. 1(d).

To investigate the stability of GDNR, molecular dynamics (MD) simulations were performed to reveal possible thermal-induced bond forming, breakage or drift processes. The simulation system was an AGDNR or ZGDNR with periodic boundary condition and the Brenner potential [29, 30] was used for C-C interactions. By the velocity Verlet algorithm and a time step of 0.2 fs, simulation was initialized at a given temperature $T$ and a thermal bath was applied, which randomly chooses an atom $i$ and replaces its velocity $\bar{v}_i^{old}$ with $\bar{v}_i^{new}$ in a time interval [31]. Here,

$$\bar{v}_i^{new} = (1-\theta)^{1/2} \bar{v}_i^{old} + \theta^{1/2} \bar{v}_i^T \qquad (i=x,y,z), \qquad (2)$$

where $\bar{v}_i^T$ is a random velocity chosen from the Maxwellian distribution. $\theta$ is a parameter controlling the strength of velocity reset, whose value was found in our previous works about C systems [32, 33] to be the most effective for temperature control.

At $T$>1700 K, the bond forming, breakage and drift frequently take place within a time $t$ less than 100 ps. As an example, Fig. 2(a) shows the snapshots in the evolution of 4-AGDNR at $T$=2200 K. In the reconstruction progress, the 4-GDNR gradually changes into an irregular structure with pentagonal, hexagonal and heptagonal rings. To evaluate the thermal reconstruction rate, an integer $N$ was used to denote the number of atoms whose bonding geometry has changed. For example, $N$ increases by 2 when a new bond forms or an old bond drifts. The corresponding inverse processes, in which $N$ decreases, may occasionally happen. The average reconstruction rate $R=\overline{\Delta N/\Delta t}$. An example for the evolution of $N$ is shown in Fig. 2(b), where $R$ is the slope of linear fitting line in the beginning stage ($t=0\sim25$ ps). At $T=1700\sim2300$ K, simulations were performed 10 times at every temperature for 2~5-AGDNR and ZGDNR to get the average value of $R$, which was found growing with increasing temperature and roughly satisfying $R\approx R_0 exp(-E_0/kT)$, where $R_0\approx3.6\times10^2$ ps$^{-1}$ and $E_0\approx1.2$ eV. At $T=300$ K, it can be estimated that $R\approx2.5\times10^{-6}$ s$^{-1}$, corresponding to $\tau=1/R\approx1.1\times10^2$ hours for once bond reconstruction event taking place. For the simulation system including 288 bonds, the bond reconstruction probability is $1/288\tau\approx0.08\%$ per day. The above result indicates that the GDNRs are stable at room temperature.

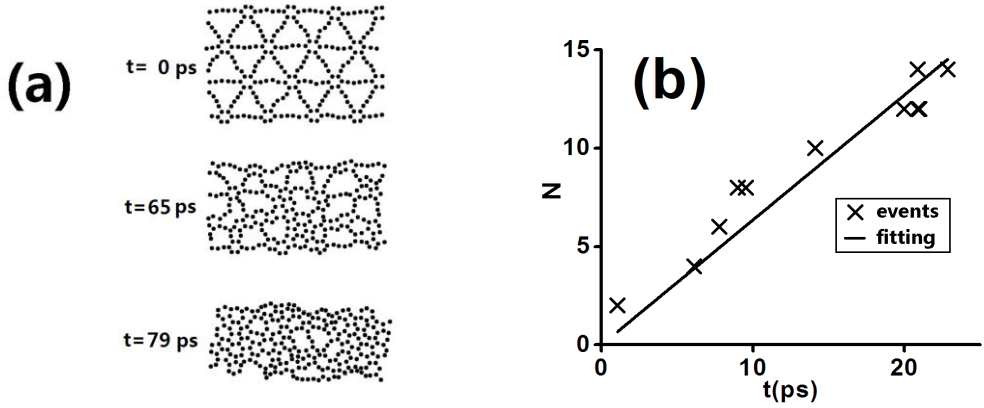

Fig. 2 The evolution of 4-AGDNR at $T=2200$ K (a), and the corresponding $N$ varying with $t$ and a linear fitting of $N$-$t$ relation (b).

3.2 *Modulation of electronic properties*

To modulate the electronic properties, adatom on GDNR surface was considered. Metal atoms may be good electron donors because of their contribution of valence electrons to the conduction band of GDNR. Furthermore, the geometry of GDNR would not be severely distorted because the bonding of metal atoms has no obvious directivity, and the band structure could be kept. By the control of chemical vapor deposition rate, the density of metal atoms on GDNR, which corresponding to the concentration of current carrier, could be artificially modulated.

Before studying the electronic properties, the thermal stability of metal atoms on GDNR surface was investigated. Geometry optimizations were performed for Au, Cu, Fe, Ni and Pt atoms on GDNR, as the model in Fig. 3(a) with the unit cell shown by the solid line. According to the result, the metal atoms all locate in the GDNR surface. Cu, Fe, Ni and Pt atom are in the angle of C triangle [Fig. 3(a)] (similar to Li [34]), while Au locating in the triangular center. The binding energy $E_b=E(\text{GDNR+Metal})-E(\text{GDNR})-E(\text{Metal})$ of Au, Cu, Fe, Ni and Pt are 3.4~7.7 eV [Table 1]. For Fe, Ni and Pt, the lengths and angles of $sp$ C-C bonds near the metal atom slightly change since the binding energy is large, while for Au and Cu the change is not obvious. For two metal atoms decorated in a same C triangle, the binding energy per atom $E_{b2}=[E(\text{GDNR+2Metal})-E(\text{GDNR})-2E(\text{Metal})]/2$ is close to $E_b$ [Table 1]. The MEP of one metal atom approaching GDNR surface was found barrierless, i.e. the barrier for one metal atom escaping from GDNR is equal to the binding energy $E_b$. Then the escaping rate at temperature $T$ could be approximately evaluated by $r=r_0 exp(-E_b/kT)$, where $r_0$ is an empirical factor corresponding to the attempt vibration frequency (generally in a magnitude of $10^{13}$ s$^{-1}$). At $T$=300 K, the escaping rate of Au is about $10^{13}$ s$^{-1}\times exp(-3.4\text{ eV}/kT)$=8$\times 10^{-45}$ s$^{-1}$, corresponding to a lifetime of $\tau=1/r\approx 4\times 10^{36}$ years. At $T$=900 K, the lifetime decreases to 300 hours. For Cu, Fe, Ni and Pt whose $E_b$ are larger than Au, the corresponding lifetimes are much longer. Then, to investigate the diffusion of metal atoms on the GDNR surface, the MEP of one metal atom migrating to the adjacent C triangle was calculated, and the diffusion barrier $E_d$ was listed in Table 1. The diffusion rate could be also estimated by $r_d\approx 10^{13}$ s$^{-1}\times exp(-E_d/kT)$, corresponding to diffusion time $\tau_d=1/r_d\approx$77 days, 38 months, $2\times 10^{43}$ years, $7\times 10^{24}$ years and $6\times 10^{12}$ years for Au, Cu, Fe, Ni and Pt, respectively. The above result indicates that at room temperature these metal atoms are very stable in GDNR surface.

To find other possible stable positions, MEP of the metal atom moving from the C triangle to the nearest hexagonal C ring were calculated. For Au, Cu, Fe, Ni and Pt, the potential energy increases along MEP and the metal atom gradually leaves the GDNR surface. No potential minima were found near the hexagonal C ring. To confirm this result, *ab initio* MD simulation was performed for these metal atoms with the same DFT scheme and Nosé thermostat. At the beginning, the metal atom was put on the hexagonal C ring. At $T=1000$ K, it migrates to the C triangle in about 2 ps. So, the positions for metal atoms in the C triangle are of the lowest potential energy.

|            | Au  | Cu  | Fe  | Ni  | Pt  |
|------------|-----|-----|-----|-----|-----|
| $E_b$ (eV)  | 3.4 | 4.6 | 7.4 | 7.7 | 5.1 |
| $E_{b2}$ (eV) | 2.6 | 4.5 | 8.0 | 7.7 | 5.0 |
| $E_d$ (eV)  | 1.2 | 1.3 | 3.3 | 2.7 | 2.0 |

Table 1 The binding energy $E_b$ and $E_{b2}$ and the diffusion barrier $E_d$ for metal atoms in the GDNR surface.

By the unit cell shown in Fig. 3(a) (192 C atoms and 1 metal atoms, corresponding to a doping ratio of 0.52%), The energy bands of 4-AGDNR decorated by Au, Cu, Fe, Ni or Pt atoms were plotted by 100×1×1 *k*-sampling with the same unit cell as the above simulation. According to the result, the bands [Fig. 3(b)] and corresponding DOS [Fig. 3(c)] of these metal-decorated 4-AGDNRs have similar feature to that of pure 4-AGDNR. For Au, the contribution of its valence electrons lifts the Fermi surface of GDNR to the bottom of conduction band and turns the GDNR into a conductor. The valence electrons of Cu have more contribution, lifting the Fermi surface into the conduction band. The Fe, Ni and Pt-decorated 4-AGDNRs are n-type semiconductors with Fermi surface below the conduction band, and impurity states were found between the valance and conduction band. The band gaps of Au, Cu, Fe and Pt-decorated 4-AGDNRs [Table 2] are in the range of 0.60~0.66 eV which are close to that of pure 4-AGDNR (0.66 eV), while Ni-decorated 4-AGDNR has a band gap of 0.76 eV. The current carrier concentrations of metal-decorated GDNRs can be evaluate by $N=n+p$. Here,

$$n = \int \frac{D(E)dE}{e^{(E-E_F)/kT}+1} \qquad (3)$$

is the electronic concentration in the bands above Fermi level and

$$p = \int D(E)(1 - \frac{1}{e^{(E-E_F)/kT}+1})dE \qquad (4)$$

is the hole concentration in the bands below Fermi level, where $D(E)$ and $E_F$ denote the DOS and Fermi energy respectively. To precisely plot the DOS near Fermi level, $E=E(k)$ of each band was fitted by polynomial and correspondingly $D(E) = \frac{dN}{dE} = \frac{Nl}{2\pi}\frac{dk}{dE}$, where $N$ the number of $k$-points and $l$ the lattice constant. Note, band lines with a same $E(k)$ but different spin are treated as two bands. At $T$=300 K, the carrier concentration of pure 4-AGDNR ($n=p$) is $N$=2.2×10$^6$ cm$^{-2}$. For Au and Cu-decorated 4-AGDNRs ($n \gg p$), $N$ are up to 2.5×10$^{11}$ cm$^{-2}$ and 8.9×10$^{11}$ cm$^{-2}$ respectively, getting close to that of graphene (~10$^{12}$ cm$^{-2}$). For Fe, Ni and Pt-decorated 4-AGDNRs, the electrons in the impurity states below Fermi surface contribute extra current carriers into the conduction band by thermal excitation, and the corresponding $N$ are 3.6×10$^7$ cm$^{-2}$, 1.6×10$^9$ cm$^{-2}$ and 1.6×10$^7$ cm$^{-2}$, respectively. To evaluate the conduction ability of single metal atom on GDNR, quantum transport calculations were performed for a piece of 4-AGDNR decorated with single metal atom sandwiched between two pure 4-AGDNR electrodes (the sketch in upper Fig. 3(d)). For pure and all of the metal-decorated 4-AGDNRs, obvious current appears when the bias voltage exceeds the band gap. In the voltage range of $V_b$=0.8~1.2 V, the currents of pure 4-AGDNR are in the range of $I$=0.6~19.0 μA. For 1.0 V, the currents of Au, Cu, Fe, Ni and Pt-decorated 4-AGDNRs are 14%, 18%, 28%, 10% and 5% larger than the pure one, respectively. This is because the impurity states of doping atoms become extra transport channels. According to above result, the conductivity of GDNR could be changed by decorating Au, Cu, Fe, Ni or Pt atoms without severely changing the energy band. Combining metal atom decorating and the band gap modulating method by transverse electric fields [35], the electronic property of GDNR could be systematically controlled.

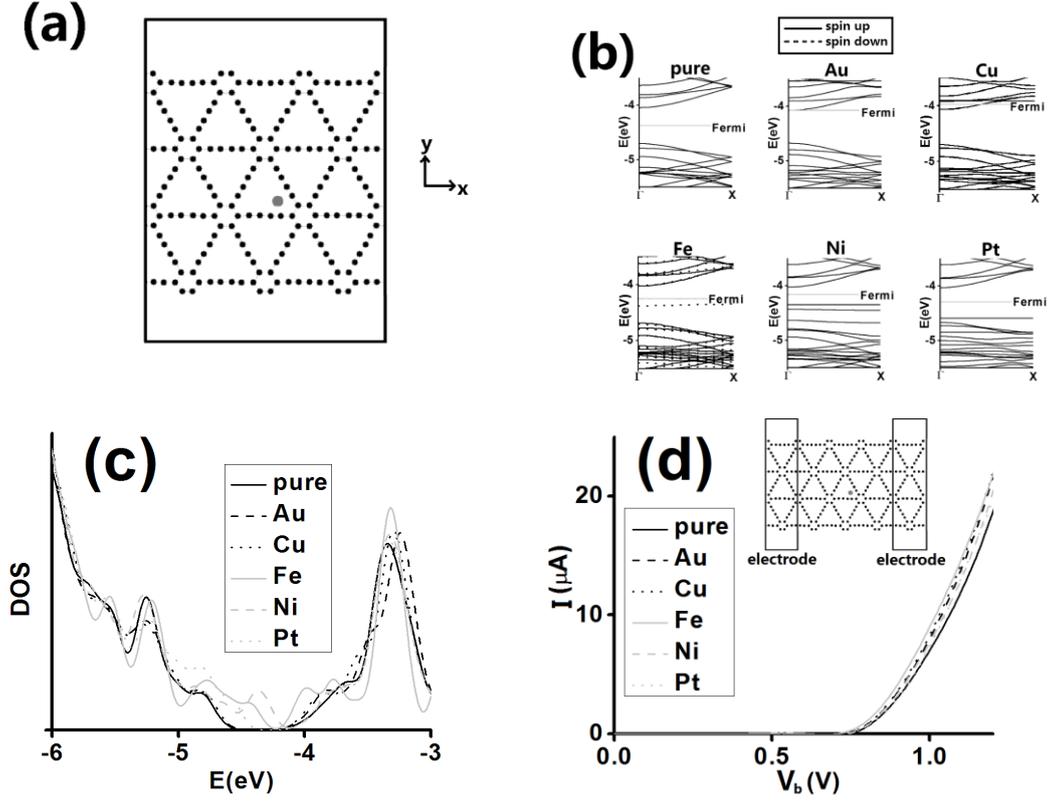

Fig. 3 The unit cell for the simulation of Au, Cu, Fe, Ni or Pt atoms (gray) on 4-AGDNR (a) and the corresponding energy bands (b) and DOS (c) for the optimized structure of pure or metal-atom-decorated 4-AGDNR. The current-voltage curve for pure and Au-, Cu-, Fe-, Ni- and Pt-decorated 4-AGDNR (d), with corresponding structure for the quantum transport calculation shown in the upper sketch.

|  | pure | Au | Cu | Fe | Ni | Pt |
| --- | --- | --- | --- | --- | --- | --- |
| band gap (eV) | 0.66 | 0.60 | 0.63 | 0.66 | 0.76 | 0.64 |

Table 2 The band gap of pure and Au-, Cu-, Fe-, Ni- and Pt-decorated 4-AGDNRs.

Since it has been reported that the electronic properties of transition metal adatom on graphene [36], graphyne and graphdiyne [37] are affected by the localization effect of $d$ electrons, it is essential to investigate this effect and verify the above results by LDA+U method. Here, calculations based on Dudarev's LDA+U formalism with $U_{\text{eff}}=U-J=5.5$ eV were performed to compare with the DFT-GGA results. Projector augmented-wave pseudopotentials are used with a cutoff energy of 400 eV. To avoid large computation quantity, the unit cell was simply taken as one 4-AGDNR unit cell decorated with one metal atom (corresponding to a doping ratio of 1.5%). According to the result, for Au, Cu, Ni and Pt, the LDA+U band gap is about 7~8% larger than the DFT-GGA band gap. However, for Fe the LDA+U band gap is 0.09 eV, which is much less than the DFT-GGA value of 0.40 eV. And by LDA+U the split of spin-up and spin-down band is much larger than by DFT-GGA.

3.3 *Metal-Semiconductor rectifier*

To evaluate possible application of GDNR as rectifier, metal-semiconductor heterojunction composed by pure GDNR and Au or Cu-decorated GDNR was considered. Quantum transport calculations were performed for the simulation system shown in Fig. 4(a), with pure 4-AGDNR and Au or Cu-decorated 4-AGDNR as left and right electrode, respectively. To avoid large computation quantity, the unit cell of right electrode was simply taken as the unit cell of 4-AGDNR decorated with one Au or Cu atom, corresponding to a doping ratio of 1.5%. In this case, the Fermi energies of Au and Cu-decorated 4-AGDNR are higher than the previous case (0.52% doping ratio) [Fig. 4(b)], and they consequently have much more carriers in the conduction band ($5.392 \times 10^{12}$ and $6.772 \times 10^{12}$ cm$^{-2}$ at 300 K, respectively). When a bias voltage $V_b>0$ is applied on the simulation system, the electrons in the conduction band of Au- or Cu-decorated 4-AGDNR are injected into the conduction band of pure 4-AGDNR. For $V_b<0$, the current $I^-$ should be smaller than $I^+$ for $V_b>0$ because the electrons in the conduction band of pure 4-AGDNR are much fewer than that of metallic Au- or Cu-decorated 4-AGDNR. However, in the range of $|V_b|$=0.2~0.6 and 0.2~0.7 V for Au- [Fig. 4(c)] and Cu-decorated 4-AGDNR [Fig. 4(d)], respectively, $I^-$ was found larger than $I^+$. This should be attributed to the electron transfer between the left and right electrode. Before applying the bias voltage, some electrons in the conduction band of the right electrode are injected into the left due to the difference between the Fermi energies. Therefore, at $V_b<0$ the transferred electrons return back and have a contribution to the current $I^-$. Since the Cu-decorated 4-AGDNR has more electrons in the conduction band than the Au-decorated 4-AGDNR, its electron transfer effect is stronger than the Au-decorated one, and the corresponding $|V_b|$ range for $\Gamma^->\Gamma^+$ is wider. At higher $|V_b|$, $\Gamma^+$ is obviously larger $\Gamma^-$. At $|V_b|=1.0$ V, $\Gamma^+$ is about 2.8 and 1.5 times of $\Gamma^-$ for the Au- and Cu-decorated 4-AGDNR, respectively.

To find the material with minimal electron transfer effect, the energy band profiles of Cr-, Mn-, Fe- and Co-decorated 4-AGDNR unit cell [Fig. 4(b)] were calculated. Among them, Fe-decorated 4-AGDNR has fewest electrons in the conduction band and the electron transfer effect should be the weakest of all. For $V_b>0$, the forward current $\Gamma^+$ monotonously increases with $V_b$, reaching 27 µA at $V_b =1.0$ V [Fig. 4(e)]. In the range of $|V_b|=0.4~0.6$ V, the reverse current $\Gamma^-$ is still larger than $\Gamma^+$. For $|V_b|>0.7$ V, $\Gamma^-$ becomes smaller than $\Gamma^+$. At $|V_b|=1.0$ V, $\Gamma^-=11$ µA and the corresponding rectification ratio $\Gamma^+/\Gamma^-=2.5$, which gets close to that of Au.

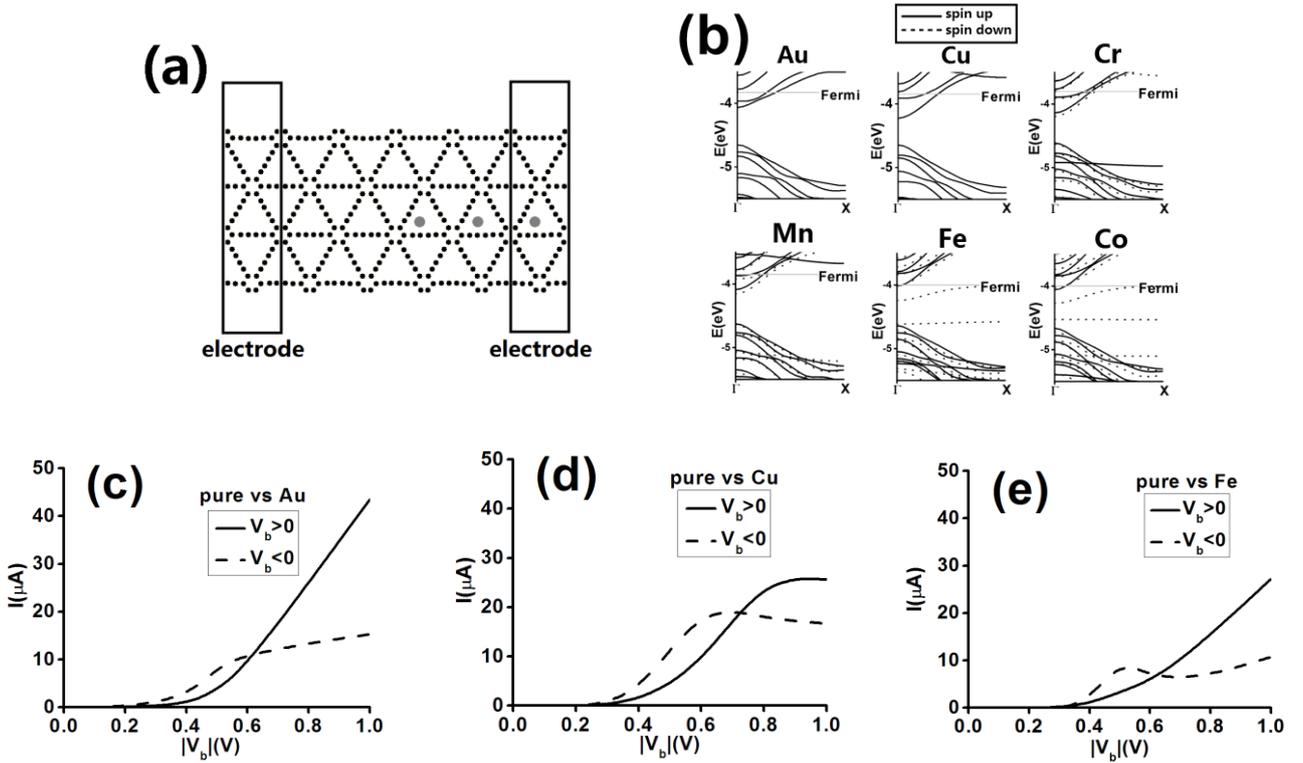

Fig. 4 (a) The simulation system of metal-semiconductor heterojunction with pure and Au- or Cu-decorated 4-AGDNR as the left and right electrode, respectively. (b) The energy band profile of 4-AGDNR decorated with one Au, Cu, Cr, Mn, Fe and Co atom per unit cell. The $I$-$V_b$ curve for the heterojunction with Au- (c), Cu- (d) or Fe-decorated 4-AGDNR (e) as the right electrode.

## 4. Conclusion

In this work, the modulation of GDNRs' electronic properties by metal adatoms was studied based on the investigation of thermal stability. By MD simulations, the average thermal bond reconstruction rate in GDNRs was obtained at high temperature, and by extrapolation to 300 K the bond reconstruction probability was predicted to be 0.08% per day, indicating that GDNRs are stable at room temperature. By *ab initio* calculation, the Au, Cu, Fe, Ni and Pt adatoms on GDNR were found very stable without severely distorting the GDNR's geometry and energy band profile. Evaluated by the adsorption energy, the thermal rate of the adatoms escaping from GDNR was predicted to be slower than $3 \times 10^{-41}$ and 0.003 atoms per hour at 300 and 900 K, respectively. At room temperature, the carrier concentration of Au- and Cu-decorated 4-AGDNR with 0.5% doping ratio are $2.5 \times 10^{11}$ cm$^{-2}$ and $8.9 \times 10^{11}$ cm$^{-2}$, respectively, getting close to graphene. The Fe, Ni and Pt-decorated 4-AGDNR were found to be n-type semiconductors with impurity states below Fermi surface. Heterojunction composed by doping Au, Cu or Fe atoms on one side of 4-AGDNR was proposed as metal-semiconductor rectifiers with a rectification ratio of 2.8, 1.5 or 2.5 at 1.0 V, while the reverse current was found larger than the forward one at small bias due to the charge transfer effect.

<p style="text-align:center">∗∗∗</p>

## Acknowledgements

This work was supported by the National Natural Science Foundation of China under Grant No. 11304239, and the Fundamental Research Funds for the Central Universities.